\documentclass[journal=langd5,manuscript=article]{achemso}
\DeclareSymbolFont{matha}{OML}{txmi}{m}{it}
\DeclareMathSymbol{\varv}{\mathord}{matha}{118}
\usepackage[version=3]{mhchem} 
\usepackage{amssymb}
\usepackage{xcolor}
\usepackage{soul,ulem}
\usepackage{physics}
\usepackage{upgreek,textgreek}

\newcommand{\blu}[1]{{\color{black}{#1}}}

\author{Russell Kajouri}
\affiliation[ifpan]
{Institute of Physics, Polish Academy of Sciences, Al. Lotnik\'ow 32/46, 02-668 Warsaw, Poland}
\author{Panagiotis E. Theodorakis}
\affiliation[ifpan]
{Institute of Physics, Polish Academy of Sciences, Al. Lotnik\'ow 32/46, 02-668 Warsaw, Poland}
\email{panos@ifpan.edu.pl}
\author{Jan \v{Z}idek}
\affiliation{Central European Institute of Technology, Brno University of Technology, Purky\v{n}ova 656/123, 612 00 Brno, Czech Republic}
\author{Andrey Milchev}
\affiliation{Bulgarian Academy of Sciences, Institute of Physical Chemistry, 1113 Sofia, Bulgaria}

   \title[Anti-Durotaxis Droplet Motion onto Gradient Brush-Substrates]
  {Anti-Durotaxis Droplet Motion onto Gradient Brush-Substrates}


\begin{document}

\begin{tocentry}

\includegraphics[width=8.25cm]{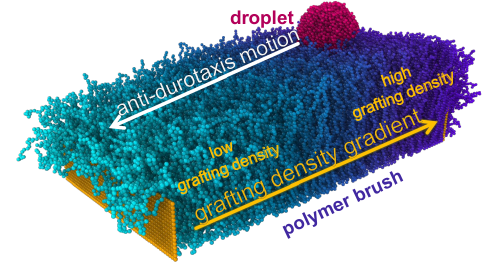}

\end{tocentry}

\begin{abstract}
Durotaxis motion is a spectacular phenomenon manifesting itself by
the autonomous motion of a nano-object between parts of a substrate
with different stiffness. This motion usually takes
place along a stiffness gradient, from softer to
stiffer parts of the substrate. Here,
we propose a new design of a polymer
brush substrate that demonstrates anti-durotaxis droplet
motion, that is droplet motion
from stiffer to softer parts of the substrate. 
By carrying out extensive molecular dynamics simulation of a 
coarse-grained model,
we find that anti-durotaxis is solely controlled by
the gradient in the grafting density of the brush and 
is favorable for fluids with a strong attraction to the
substrate (low surface energy). 
The driving force of the anti-durotaxial 
motion is the minimization of the droplet--substrate
interfacial energy, which is attributed to
the penetration of the droplet into the brush.
Thus, we anticipate that the proposed substrate design offers 
new understanding and
possibilities in the area of autonomous motion
of droplets for applications in microfluidics, 
energy conservation, and biology.
\end{abstract}

\vspace{0.7in}

\section{INTRODUCTION}
The spontaneous motion of nano-objects onto substrates
has attracted much interest,
due to its potential impact in applications, such 
as microfluidics, microfabrication, coatings, energy conversion, and 
biology.\cite{Srinivasarao2001,Chaudhury1992,Wong2011,Lagubeau2011,Prakash2008,Darhuber2005, Yao2012, Li2018, Becton2016,vandenHeuvel2007,DuChez2019,Khang2015}
In this kind of autonomous motion, the direction of motion
can also be controlled, steered by gradient changes in a
substrate property that can be ``sensed'' by the nano-object.
More specifically, such property can be
the stiffness of the substrate,
which can enable and sustain
the motion of the nano-object 
in a specific direction. 
A characteristic example here is the
motion of cells on tissues, known as durotaxis.\cite{DuChez2019,Khang2015,Lo2000,Pham2016,Lazopoulos2008}
Apart from biological systems, however, durotaxis has
also been realised in the case of a spectrum of different nano-objects 
(\textit{e.g.} fluids),
both in theoretical and simulation
models,\cite{Theodorakis2017,Chang2015,Becton2014,Barnard2015,Palaia2021,Tamim2021,Bardall2020,Kajouri2023} as
well as in experiments.\cite{Style2013}

An important aspect of the durotaxis motion is that
the nano-object can sustain the motion without
external energy supply. However, apart from stiffness gradients,
such motion can also be caused by specific substrate
patterns. Characteristic examples here
are rugotaxis, where the motion of fluids is provoked
by a gradient in the wavelength that characterizes
wavy substrates\cite{Theodorakis2022,Hiltl2016}, and
curvotaxis driven by curved protein 
complexes at the cell\cite{Sadhu2023}.
Other possibilities may include
the transport of small condensate droplets on 
asymmetric pillars\cite{Feng2020}, three-dimensional
capillary ratchets\cite{Feng2021}, or taking advantage of 
pinning and depinning effects at
the three-phase contact line\cite{Theodorakis2021}.
Moreover, in the case of capillary ratchets, 
the motion can take place along or against the gradient
depending on the surface tension of the fluid.\cite{Feng2021}
Wettability gradients have also been exploited to 
steer the motion of fluids\cite{Pismen2006,Wu2017,Sun2023},
while the long-range transport of fluids
can be realized by using electrostatic
\cite{Sun2019,Jin2022} or triboelectric charges\cite{Xu2022}.
In contrast, the use of external fields, such as in
the case of electrotaxis\cite{Jin2023}, requires
energy supply by an external source,
as, also, in the case of thermotaxis to sustain
a temperature gradient\cite{Zhang2022}.
Other options requiring external sources may include the use of
electrical current \cite{Dundas2009,Regan2004,Zhao2010,Kudernac2011},
charge \cite{Shklyaev2013,Fennimore2003,Bailey2008},
or even simple stretching\cite{Huang2014},
as well as chemically driven droplets \cite{Santos1995,Lee2002},
droplets on vibrated substrates \cite{Daniel2002,Brunet2007,Brunet2009,Kwon2022}
or wettability ratchets \cite{Buguin2002,Thiele2010,Noblin2009,Ni2022}.

In our previous studies, we have investigated various
substrate designs that can cause and sustain the 
motion of liquid droplets,\cite{Theodorakis2017,Theodorakis2022,Kajouri2023}
which were mainly motivated by relevant experiments.\cite{Style2013,Hiltl2016}
In particular, in the context of durotaxis droplet motion, 
a new design based on brush polymer substrates
was proposed, where the
stiffness gradient was imposed by varying the chain stiffness of 
the grafted polymers along the gradient
for a given grafting density.\cite{Kajouri2023} 
In this case, it has been found
that the grafting density of polymer chains and the droplet
adhesion to the brush are the key parameters
determining whether the motion will be realised, as well as
its efficiency. In particular, it has been found that moderate
values of both will promote the droplet motion.
Surprisingly, the stiffness gradient
itself, albeit necessary for the durotaxis motion, turned out
to be irrelevant for determining the efficiency of the motion in terms of the average velocity of the droplet. 
Importantly, the direction of the droplet motion was
in the same direction as the stiffness gradient,
\textit{i.e.}, from softer to stiffer parts of the 
substrate. In effect, this translates into a varying
substrate roughness, which drives the droplet motion.
In contrast, in a specific experiment,\cite{Style2013}
the direction of motion
of droplets on a soft, silicon-gel substrate has been from 
the stiffer toward the softer parts of the substrate for
\(\mathrm{\text\textmu m}\) scale droplets, which is well
below the capillary length scale ($\sim$2.5~mm) in the 
case of water droplets. 
Although for this reason gravity seems not to play
an important role, durotaxis has been more efficient 
in the case of larger droplets.\cite{Style2013}
While biological systems\cite{DuChez2019,Khang2015,Lo2000,Pham2016,Lazopoulos2008}
and simulation models\cite{Theodorakis2017,Kajouri2023,Palaia2021,Chang2015} 
have been thus far only able to demonstrate the droplet motion 
in the direction of the stiffness gradient, that is from 
softer to stiffer parts of a substrate, to the best of our
knowledge, there is currently no \textit{in silico} substrate design 
that has demonstrated droplet motion in the opposite
direction of the stiffness gradient, namely from the
stiffer toward the softer parts of the substrate.

Motivated by relevant experiments\cite{Style2013}
and previous experience with gradient brush substrates,\cite{Kajouri2023}
we consider a polymer brush that
can initiate and sustain the droplet motion toward
the softer parts of the substrate. 
Here, we will refer to this kind of motion as
``anti-durotaxis'' in order to underline the fact
that the droplet moves in the opposite direction
with respect to a positive stiffness gradient. 
In this new design of brush substrate, the stiffness gradient is 
implemented by the gradual change in the 
grafting density of fully flexible polymer chains. 
By using extensive molecular dynamics (MD)
simulations of a coarse-grained (CG) model,
we explore the key parameters of the system,
such as the gradient in the grafting density,
the droplet attraction to the substrate,
the droplet size, and the viscosity. 
Our method also provides the molecular-scale
resolution required to explore the underlying
mechanism of the anti-durotaxis motion. In this way,
our study casts further light on
the self-sustained motion of droplets onto
brush gradient substrates, and, as a result,
unravels new possibilities in nanoscale science and
technology\cite{Barnard2015} for various 
medicine and engineering applications.\cite{Barthlott2016,Khang2015}
Moreover, brush substrates share structural characteristics with
various biological surfaces that can expel various
exogenous substances from their structure,\cite{Badr2022}
such as the mucus layer from airway epithelia.\cite{Button2012}
Also, in the context of regenerative medicine,\cite{Khang2015}
the concept of gradient substrates plays an 
important role for applications in this area,
for example, in drug transport within the body.
Hence, we anticipate that our study has a broader impact
beyond engineering applications.
Moreover, since the design of the brush substrate
only depends on the variation (gradient) of the grafting density,
as compared to a previous design that the chain 
stiffness and possibly the chemistry of the chains
had to be varied to tune the substrate properties\cite{Kajouri2023},
might suggest that 
the brush system investigated here might hold
greater hope for experiments, thus offering further 
possibilities for relevant applications.
In the following,
we provide details of the system, simulation model
and methodology. Then, we will present and discuss 
the obtained results, while  we
will draw the conclusions resulting from
our investigations in the final section.

\section{MATERIALS AND METHODS}
\begin{figure}[bt!]
\centering
 \includegraphics[width=\columnwidth]{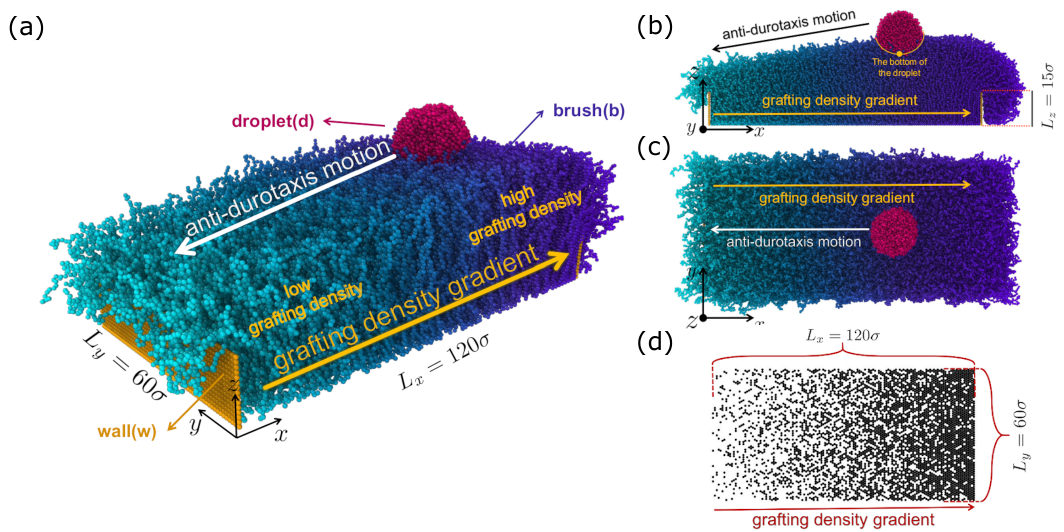}
\caption{\label{fig:1} 
(a) Typical initial configuration of the system.
The droplet is placed on the substrate side
with the highest grafting density,
which is here $\sigma_{\rm g,h}=0.6~\sigma^{-2}$, while 
on the other side the lowest grafting density is 
$\sigma_{\rm g,l}=0.1~\sigma^{-2}$. 
The length of the system in the
gradient direction, $x$, is $L_{\rm x} = 120~\sigma$, 
with the grafting density gradient defined as 
$G=(\sigma_{\rm g,h}-\sigma_{\rm g,l})/L_{\rm x}$.
Droplet and substrate polymer-chains consist of 
fully flexible chains of length $N_{\rm d}=10$ 
and $N_{\rm b}=50$ beads, respectively, while the 
total size of the droplet is $N=4000$ beads in this case.
Also, the strength of the interaction between the
droplet and substrate beads is
$\varepsilon_{\rm db}=0.9~\epsilon$. 
This particular system has shown the most
efficient
anti-durotaxial  motion 
(in terms of average droplet speed) among all of the cases
considered in our study.
(b) Side view of the same system after 
the droplet has moved a certain distance
from its starting point. An $x-z$ cross-section
passing through the center of mass of the droplet
is shown to highlight the penetration of the
substrate by the droplet.
(c) Top view of the same configuration.
(d) The distribution of the grafting sites
of the brush polymer chains on the
bottom, solid plane of immobile beads
with a honeycomb geometry is shown. 
At the right boundary, chains are randomly grafted 
in the vertical ($y$) direction with probability
$\sigma_{\rm g,h}=0.6~\sigma^{-2}$, while 
$\sigma_{\rm g,l}=0.1~\sigma^{-2}$ at the 
left most boundary.
\blu{The snapshot of the system was obtained using 
Ovito software.\cite{Stukowski2010} } 
}
\end{figure}

The system setup is illustrated in Fig.~\ref{fig:1}. 
It consists of a brush substrate and a droplet.
The brush has a gradient 
in the grafting density of the polymer chains
that are tethered to a bottom wall of immobile
beads with hexagonal (honeycomb) symmetry.
Vertical walls of immobile beads at the two ends
of the substrate that are parallel to the $y-z$ plane
with height $L_{\rm w} = 15~\sigma$ 
($\sigma$ is the length unit) are also present
to support the structure of the brush in the
direction of the gradient,
\blu{extending thus the total area of undistorted grafting density gradient closer to the brush boundaries in $x$ direction.}.
The droplet is placed on the
side with the highest grafting density
at a distance $30~\sigma$ between the center of mass 
of the droplet and the side wall, 
as shown in Fig.~\ref{fig:1}. 
After examining a range of different
scenarios, we have determined that in the current
substrate design the droplet motion will take
place from higher grafting density areas,
which implies a higher substrate stiffness, toward
areas of lower grafting density. 
In the direction of the stiffness gradient ($x$ direction), 
the length of the substrate
is $L_x = 120~\sigma$, while $L_y = 60~\sigma$ is the substrate
length in the $y$ direction, as shown in Fig.~\ref{fig:1}.
Periodic boundary conditions are considered
in all Cartesian directions. In particular,
in the $y$ direction, the boundaries of the simulation box
coincide with those of the substrate, while in the
$x$ and $z$ directions the size of the simulation box is large enough
to prevent the interaction of beads from opposite
boundaries. In view of the sufficiently
larger simulation
box than the substrate in the direction of the gradient ($x$ 
direction),
the presence of the side walls is required in order
to maintain the structure of the brush in this direction.

The standard bead-spring model\cite{Kremer1990} was
employed for the molecular dynamics simulations. 
According to this model,
the interactions between the different components of 
the system, 
\textit{i.e.}, the drop (d), the brush (b), and the wall (w) beads,
are expressed through the Lennard-Jones (LJ) potential,
which reads
\begin{equation}\label{eq:LJpotential}
U_{\rm LJ}(r) = 4\varepsilon_{\rm ij} \left[  \left(\frac{\sigma_{\rm  ij}}{r}
\right)^{12} - \left(\frac{\sigma_{\rm ij}}{r}  \right)^{6}    \right],
\end{equation}
where $r$ is the distance between any pair of beads in
the system.
The ${\rm i}$ and ${\rm j}$ indices in Eq.~\ref{eq:LJpotential}
correspond to the bead type (d, b, w). 
The size of all of the beads is set to $\sigma_{\rm ij} = \sigma$.
Moreover, the LJ potential is cut and shifted at a 
specific distance (cutoff), which for the 
interactions between the droplet beads or between the
droplet and the brush-polymer beads is $r_{\rm c}=2.5~\sigma$,
while for all other interactions
an athermal model is used
with the cutoff set to the minimum of the LJ potential,
namely $r_{\rm c}=2^{1/6}~\sigma$.
The potential well of the attractive interactions
between the droplet beads
is $\varepsilon_{\rm dd}=1.5~\epsilon$, while 
different choices for $\varepsilon_{\rm db}$ are considered 
for the interaction strength between the droplet and the
substrate.\cite{Theodorakis2011}
In particular, larger values of $\varepsilon_{\rm db}$
would correspond to fluids with smaller surface
energy, while smaller values of $\varepsilon_{\rm db}$
would be suitable to model fluids with larger
surface energy, \textit{i.e.}, fluids with a lower
tendency of wetting a substrate.
Here, $\epsilon$ is the energy unit and 
the range $\varepsilon_{\rm db}=0.1-1.2~\epsilon$ is
chosen to conduct our investigations, which allows 
us to capture all possible scenarios for the
droplet for the specific substrate design.
\blu{For all other interactions, such as those between
the droplet and the walls or the brush polymers and the
walls, the interaction strength is set equal to $\epsilon$, while,
as mentioned above, the model is anyway athermal
for these repulsive interactions.}

The size of the droplet can vary,
ranging from $4\times10^3$ 
to $16\times10^3$ beads in our study.
These beads are parts of fully flexible,
linear polymer chains. While in our investigations the length
of the droplet chains is $N_{\rm d}=10$ beads 
throughout the different simulation cases, which ensures
that there are no evaporation effects
and the vapor pressure is hence sufficiently 
low\cite{Tretyakov2014,Kajouri2023},
chain lengths of up to $80$ beads were
also considered for particular cases to explore
the effect of viscosity on the anti-durotaxis 
motion.\cite{Kajouri2023,Theodorakis2017}
While the exact value of the viscosity is not
of importance here, and while
the scaling of the viscosity with the 
chain length in the model seems not 
to have been completely settled in the literature,
we can however note that the viscosity is 
expected to grow with the chain length. 
Moreover, a linear growth with the Kuhn length
for Brownian models of melts has been observed
on the basis of Rouse dynamics\cite{Everaers2020}
or a power-law relation $[\eta]=KM^{\alpha}$
according to Mark--Houwink--Sakurada, 
with $M$ being the molecular weight of the polymer
and $K$ is determined
by the intrinsic properties of the used
polymer, while $\alpha=0.8$ for a good-solvent
system\citep{rubinstein}.
Hence, we expect that in melt conditions the viscosity 
shall grow with a power-law exponent lower than unity
for the lengths considered here, while entanglement effects
are expected to also play a role for longer polymer 
chains. Having said that, the viscosity depends
on the particular conditions (\textit{e.g.}, solvent 
conditions) when measured, but it is generally
expected to increase with the chain length $N_{\rm d}$
of the chains.
The length of the tethered brush polymers, which are
of linear molecular architecture, is 
$N_{\rm b}$, and remains the same in all of
our \textit{in silico} experiments. 
Based on preliminary tests, 
we found that this choice allows us to remove any 
significant dependence of
the results on the choice of the brush length, $N_{\rm b}$, 
which is also in line with our previous experience with brush
substrates.\cite{Kajouri2023}

To tether the beads together in each polymer chain
of the droplet or the brush, 
the finitely extensible nonlinear elastic (FENE) bond
potential was applied for consecutive pairs of beads
in each chain, which mathematically is expressed
as follows:
\begin{equation}\label{eq:KG}
 U_{\rm FENE}(r) = -0.5 K_{\rm FENE} R_{\rm 0}^2 \ln \left[ 1 - \left(\frac{r}{R_{\rm 0}} \right)^2  \right].
\end{equation}
In the above relation, $r$ is the distance between
the beads pair and $R_{\rm 0}=1.5~\sigma$,
which determines the highest possible extension of the
bond. $K_{\rm FENE} = 30~\epsilon/\sigma^2$
is an elastic constant.

The stiffness gradient is realized
by varying the grafting density of the polymer chains
in the $x$ direction. 
By considering a hexagonal symmetry of
the possible grafting sites on the bottom substrate
of immobile beads,
chains are randomly grafted to satisfy the required
grafting density in the $x$ direction (Fig.~\ref{fig:1}d).
In particular, the substrate
is divided into bins with length $3~\sigma$ along the
$x$ direction. At the one end of the substrate we impose
the highest grafting density $\sigma_{\rm g,h}$ in units
of $\sigma^{-2}$, while at the other end the lowest grafting
density is denoted as
$\sigma_{\rm g,l}$. A linear gradient 
in the grafting density between the two 
opposite ends is considered, namely
$G = (\sigma_{\rm g,h} - \sigma_{\rm g,l}) / L_x$. 
After conducting extensive 
preliminary investigations with different choices 
for $\sigma_{\rm g,h}$ and $\sigma_{\rm g,l}$, 
we have concluded that setting $\sigma_{\rm g,l}=0.1~\sigma^{-2}$
gives us the highest number of successful 
anti-durotaxis cases,
that is, cases that the droplet is able to cross 
against the gradient the
whole distance from the one end of the substrate to the 
other. This comes as an advantage when 
investigating the influence of various parameters,
such as the droplet size, viscosity or adhesion strength,
since more data of successful anti-durotaxis cases
can be acquired for the analysis.
Finally, the range $\sigma_{\rm g,h}=0.4-0.9~\sigma^{-2}$
was considered for the highest grafting density
of the systems.
Lower values of $\sigma_{\rm g,h}$ would result in small
gradients, which would prevent the anti-durotaxis motion.
The choice $\sigma_{\rm g,h} \geq 1.0~\sigma^{-2}$ already
imposes large steric effects between the polymer chains
and their close-packing, which would also practically imply
small stiffness gradients experienced by the droplet.
We have also examined systems with stiff polymer
chains for the brush and concluded that 
anti-durotaxis motion was more efficient in the case
of fully flexible brush chains (no angle potential along
each brush chain).
\blu{This is mainly due to a smaller range of possibilities
for the stiffness gradient of the substrate when individual
polymer chains
are stiffer. In addition, stiff chains prevent the 
penetration of the substrate by the droplet,
which, as will become apparent later, hinders
the anti-durotaxis motion.}
Hence, all of the results presented
in this study will refer to brush substrates with
fully flexible polymer chains.

To control the temperature of the system, $T=\epsilon/k_B$
($k_B$ is Boltzmann's constant),
the Langevin thermostat was used\cite{Theodorakis2011}. 
Moreover, the coordinates of each bead
$\{\textbf{r}_i(t)\}$ evolve in time by integrating the Langevin 
equation by means of
the HOOMD-Blue package,\cite{hoomd-blue} which mathematically reads:
\begin{equation}\label{Eq3}
 {\rm m}\frac{d^{2}\textbf{r}_{\rm i}}{dt^{2}}=-\nabla U_{\rm i}- 
\gamma \frac{d\textbf{r}_{\rm i}}{dt}+\Gamma_{\rm i} (t).
\end{equation}
Here, $\rm m$ is the mass of the beads, which is equal to $m$, 
with $m$ being the mass unit.
$t$ denotes the time, $U_{\rm i}$ is the total potential acting on the $\rm i$-th
bead, $\gamma$ is the friction coefficient, and ${\Gamma}_{\rm i}(t)$ is the random force. As
is well-known, $\gamma$ and $\Gamma$ are related by the usual fluctuation--dissipation relation
\begin{equation}\label{Eq4}
<\Gamma_{\rm i}(t) \cdot \Gamma_{\rm j}(t^{'})>=6k_{B}T\gamma\delta_{\rm ij}\delta(t-t^{'}).
\end{equation}
Following previous work\cite{Grest1993,Theodorakis2011,Murat1991}, 
the friction coefficient was
chosen as $\gamma=0.1~\tau^{-1}$. Equation~\ref{Eq3} 
was integrated using an integration time step
of $\Delta t=0.005~\tau$, where
the time unit is $\tau=(m\sigma^{2}/\epsilon)^{1/2}$. 
Given the choice of the Langevin thermostat, the
simulations are in practice carried out in the
canonical statistical ensemble. 
For each set of parameters, we obtained an
ensemble of thirteen statistically independent trajectories
by changing the
initial velocities assigned to each particle
in order to acquire reliable statistics
for the analysis of the results. The length of 
each trajectory was $10^8$~MD integration-steps,
unless the droplet managed to reach the 
side of the substrate
with the lowest grafting density before
the set maximum simulation time. 
In this case, the simulation is terminated and the
particular case is considered as a successful 
anti-durotaxis case.

\section{RESULTS AND DISCUSSION}

\begin{figure}[bt!]
\centering
 \includegraphics[width=\columnwidth]{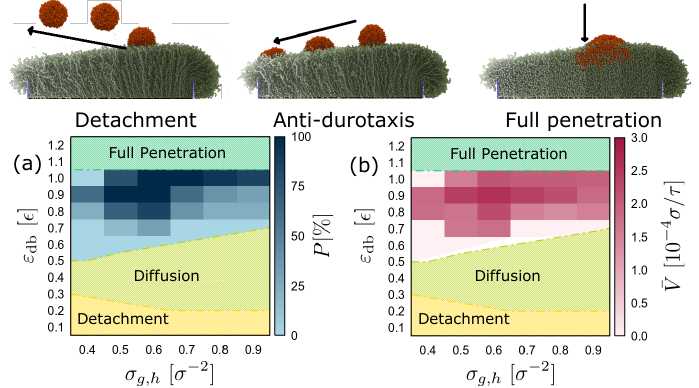}
\caption{\label{fig:2} 
(a) Regime map indicating the
probability, $P$ (color scale),
that a droplet will cover the full distance
over the substrate in the $x$ direction from the
stiffest (highest grafting density) to the softest
(lowest grafting density) part of the substrate
(successful anti-durotaxis case) 
for different values of the droplet--substrate
attraction, $\varepsilon_{\rm db}$, and the 
highest grafting density, $\sigma_{\rm g,h}$.
The lowest grafting density always has the same
value, namely $\sigma_{\rm g,l}=0.1~\sigma^{-2}$.
The probability, $P$, 
is calculated from an ensemble of 
thirteen independent simulations
for each set of parameters,
$\varepsilon_{\rm db}$ and $\sigma_{\rm g,h}$. 
The regimes that the droplet immediately penetrates
into the brush and is not able to further move 
(``Full Penetration''), detaches from the brush
due to the weak droplet--brush attraction (``Detachment''),
or carries out a random walk onto the substrate (``Diffusion'')
are also shown with a different color.
(b) The color map indicates the average
velocity of the droplet, $\bar{\varv}=L'_x/t$, for the successful 
durotaxis cases, where $t$ is the time that the droplet
needs to cross the full length of the brush substrate
in the $x$ direction, and $L'_x$ is the actual
distance covered by the center-of-mass of the droplet
for each successful case. 
$N=4000$, $N_{\rm d}=10$, and $N_{\rm b}=50$ beads.
Snapshots on top of the plot indicate examples of 
detachment, anti-durotaxis, and penetration.
For the sake of providing a perspective of the various
processes, in the former two cases,
a time sequence of droplet snapshots 
at different times are shown (during this time sequence, the
configuration of the brush substrate also changes with time, but
a single visualization of the brush substrate is shown
here only for demonstration),
while in the latter case
a single snapshot of a cross-section view
for the droplet in the brush is illustrated. 
\blu{The snapshot of the system was obtained using 
Ovito software.\cite{Stukowski2010} } 
}
\end{figure}

In the case of the durotaxis motion onto a brush,
we have determined that the key parameters of the substrate design
for a droplet of a given size are
the strength of droplet attraction to the substrate
and the grafting density.\cite{Kajouri2023} 
Surprisingly, we have also found that the stiffness gradient,
albeit necessary to cause the durotaxis motion of the 
droplet, was not \textit{per se} key for determining the efficiency
of the motion in terms of the time the droplet had required
to transverse the full length of the substrate
in the simulation.
Following a similar protocol in the case of the
anti-durotaxis phenomenon, which includes
an extensive exploration of the parameter space relevant
for the new brush design, we find that the 
grafting density at the soft end of the substrate
should be small, namely $\sigma_{\rm g,l} = 0.1~\sigma^{-2}$. 
In addition, by investigating a relevant range of
systems with different chain stiffness,
we have determined that the most efficient
anti-dutoraxis motion occurs when the 
brush chains are fully flexible. 
Henceforth, all of the presented results will therefore
consider $\sigma_{\rm g,l} = 0.1~\sigma^{-2}$ and fully flexible
polymer chains. By taking this into account, 
the key parameters determining the efficiency of 
the \blu{anti-durotaxis} motion are the stiffness gradient, which is defined by the 
value of the grafting density at the stiffest (highest grafting density)
end of the substrate with grafting density, $\sigma_{\rm g,h}$, 
and the droplet--brush attraction strength, 
which is controlled by the LJ parameter $\varepsilon_{\rm db}$.

Figure~\ref{fig:2} summarizes the results of the
simulations in the form of regime maps with
$\sigma_{\rm g,h}$ and $\varepsilon_{\rm db}$
as parameters. Different scenarios for the 
behavior of the droplet are possible during the
simulations, \textit{i.e}, its detachment from, 
or its full penetration into the brush, 
a diffusion- or random-walk-like motion, 
and, finally, anti-durotaxis motion. 
In particular, the detachment of the droplet
takes place for low attraction strength
$\varepsilon_{\rm db}$. Due to the thermal 
fluctuations the droplet is not able to 
permanently stick to the substrate, which
is more probable to happen when the grafting density
is lower, due to a smaller number of
droplet--substrate bead-pair interactions.
Moreover, we observe that the dependence on the 
grafting density disappears beyond a 
certain threshold, namely
$\sigma_{\rm g,h} \gtrsim 0.6~\sigma^{-2}$, 
which might imply that the brush density 
at the interface has reached an adequately
high value.
In contrast, full penetration takes place when
$\varepsilon_{\rm db} \geq \epsilon$, 
irrespective of the grafting density,
which suggests that full penetration
\blu{be} determined by the strong microscopic interactions
between the substrate and the droplet beads.
In this case, the
droplet shape is significantly distorted by the brush chains 
to the extent that brush chains penetrate through
the droplet chains leading to a ``non-coherent'' droplet
(see snapshot of Fig.~\ref{fig:2}).
In the third scenario, the droplet 
exhibits a diffusion-like (random-like) motion onto
the substrate, which takes place for
low $\varepsilon_{\rm db}$ values. 
This refers to unsuccessful anti-durotaxis cases
that are not able to cross the full
length of the substrate in the $x$ direction, that is in this case
the driving force of the motion is very weak.
Moreover, the range of $\varepsilon_{\rm db}$ that such
a behavior is observed
increases when the grafting density $\sigma_{\rm g,h}$
increases, which already suggests that denser brushes would
hinder the possibility of anti-durotaxis motion.
In contrast, the decreasing grafting density promotes
the motion, and the specific reasons for this
will become clearer later, when we will discuss the
underlying mechanism of the anti-durotaxis phenomenon.

\begin{figure}[bt!]
    \centering
    \includegraphics[width=\textwidth]{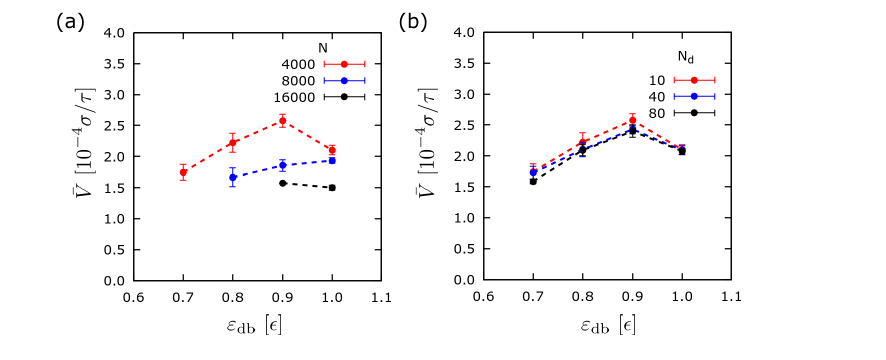}
    \caption{ \label{fig:3} (a) Average velocity of the
    droplet, $\overline{\varv}$, as a function  of the attraction strength,
    $\varepsilon_{\rm db}$, for successful anti-durotaxis cases
    for different droplet size, $N$, as indicated. 
    \blu{The average
velocity of the droplet is, $\bar{\varv}=L'_x/t$, where $t$ is the time that the droplet
needs to cross the full length of the brush substrate
in the $x$ direction, and $L'_x$ is the actual
distance covered by the center-of-mass of the droplet
for each successful anti-durotaxis case.} 
    $\sigma_{\rm g, h} = 0.6~\sigma^{-2}$,
    $\sigma_{\rm g, l} = 0.1~\sigma^{-2}$, $N_{\rm d}=10$
    and $N_{\rm b}=50$ beads. 
    (b) The average velocity for different chain
    lengths $N_{\rm d}$ of the droplet, as indicated.
    $\sigma_{\rm g, h} = 0.6~\sigma^{-2}$,
    $\sigma_{\rm g, l} = 0.1~\sigma^{-2}$, $N_{\rm d}=10$,
     $N_{\rm b}=50$, and
     $N=4000$ beads.
 }
\end{figure}

Anti-durotaxis motion is observed for 
$\varepsilon_{\rm db} \geq 0.7~\epsilon$, 
which suggests that a relatively high wettability
of the brush by the droplet favor the motion. 
Moreover, we were able to find anti-durotaxis
for the whole range of the grafting density
gradients shown in Fig.~\ref{fig:2}a, but 
a moderate choice of the highest grafting density,
in the range 
$0.5~\sigma^{-2} \leq \sigma_{\rm g, h} \leq 0.7~\sigma^{-2}$ 
(the optimum choice of
the grafting density on the soft end 
is $\sigma_{\rm g, l} = 0.1~\sigma^{-2}$),
generally presents a high certainty for successful anti-durotaxis
motion with high efficiency in terms of the time required
to move from the one end to the other end of the substrate.
By examining more carefully this time and
based on this calculate the average velocity
of the droplets during the droplet motion (Fig.~\ref{fig:2}b), 
we find that the optimum choice is 
$\sigma_{\rm g, h} \geq 0.6~\sigma^{-2}$ and
$\varepsilon_{\rm db}=0.9~\epsilon$ when the
droplet size is $N=4000$ beads (Fig.~\ref{fig:3}).

The results of Fig.~\ref{fig:3}a
indicate that increasing $\varepsilon_{\rm db}$
initially renders the droplet motion more
efficient in terms of the average droplet velocity, 
but further increase, namely $\varepsilon_{\rm db}=\epsilon$, 
leads to a smaller velocity. The latter is attributed
to the partial penetration of the brush
chains into the droplet due to the strong
attraction, while for
$\varepsilon_{\rm db}>\epsilon$ we enter 
the ``full penetration'' regime (Fig.~\ref{fig:2}) 
and the droplet motion cannot take place.
As the droplet size increases, average velocities
are evidently lower than those reported in the case
of droplet size with $N=4000$ beads.
However, it would be
difficult to identify any trends with $\varepsilon_{\rm db}$
for the larger droplets, since the increase of the
droplet size leads to a smaller number of 
successful anti-durotaxis cases. 
For example, while the choice 
$\varepsilon_{\rm db}=0.7~\epsilon$ yields
successful anti-durotaxis motion, this choice
does not show any success for droplets with
size of either $N=8000$ or $N=16000$ beads. 
A final notice concerns the behavior as
the droplet viscosity increases (Fig.~\ref{fig:3}b).
Our results indicate that its effect is rather 
minor, since a similar behavior is observed for the
droplets with chains of different $N_{\rm d}$ 
for different choices of attraction strength
$\varepsilon_{\rm db}$. However,
less viscous droplets (\textit{e.g.}, $N_{\rm d}=10$ beads) appear
to slightly favor the anti-durotaxis motion, but
differences are rather within a statistical
uncertainty. Finally, a decrease in the velocity
for $\varepsilon_{\rm db}=\epsilon$ is observed,
which is again attributed to the partial penetration
of the brush chains into the droplet. Interestingly, this effect
appears to be independent of the chain length and
fully attributed to the microscopic interactions
between the substrate and the droplet beads.

\begin{figure}[bt!]
\centering
 \includegraphics[width=\textwidth]{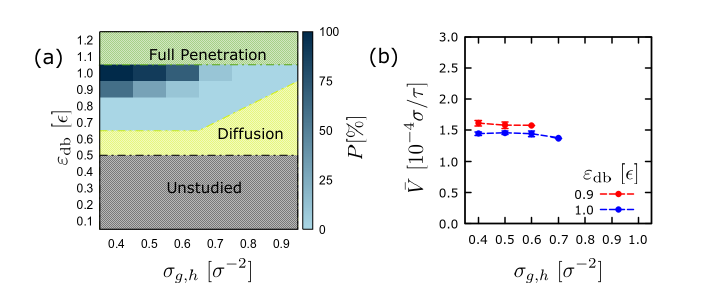}
\caption{\label{fig:4} (a) Same as in Fig.~\ref{fig:2}a,
but results are shown for cases that the droplet
consists of $N=16000$ beads in total.
(b) Average velocity, $\overline{\varv}$, for different
gradients as defined through $\sigma_{\rm g, h}$
for two different values of $\varepsilon_{\rm db}$ 
for which successful anti-durotaxis motion
takes place. $\sigma_{\rm g, l}=0.1~\sigma^{-2}$,
$N_{\rm d}=10$ and $N_{\rm b}=50$ beads.
}
\end{figure}

To carefully investigate the effect of the droplet
size, we conducted extensive \textit{in silico}
experiments with
a larger droplet, namely $N=16000$ beads.
Figure~\ref{fig:4} presents results for 
this droplet size, which generally
confirm the aforementioned observations. 
Furthermore, the regime map, which 
is again based on
the parameters $\varepsilon_{\rm db}$ and
$\sigma_{\rm g, h}$, shows that only
a few combinations of these parameters 
were able to lead to successful anti-durotaxis motion,
and, moreover, among them, only the 
set of parameters $\varepsilon_{\rm db}=\epsilon$
and $\sigma_{\rm g, h}=0.4~\sigma^{-2}$ was 
successful in 100\% of our \textit{in silico}
experiments. In addition, we observe that
smaller values of $\sigma_{\rm g, h}$ 
favor the anti-durotaxis motion in the case
of the droplet with $N=16000$ beads, 
in comparison with the case of $N=4000$ beads (Fig.~\ref{fig:2}).
This suggests that larger droplets require
a much softer substrate in order to penetrate
into the brush during the anti-durotaxis motion.
Also, for both larger and smaller droplets,
we observe that anti-durotaxis is favored by
larger values of $\varepsilon_{\rm db}$.
By examining the velocity for the successful 
cases of $\varepsilon_{\rm db}=0.9~\epsilon$
and $\varepsilon_{\rm db}=~\epsilon$ (Fig.~\ref{fig:4}b), 
we see that the average velocity remains independent
of the choice of $\varepsilon_{\rm db}$ and 
that the motion is slightly more efficient 
when $\varepsilon_{\rm db}=0.9~\epsilon$.
Hence, only a narrow range
of $\varepsilon_{\rm db}$ values can lead 
to successful anti-durotaxis motion when the
droplet size is $N=16000$ beads. Also, the range
of $\sigma_{\rm g, h}$ for successful motion
is more limited in the case of larger droplets. 
In summary, our data hint that
more space is needed within the brush to 
accommodate the droplet. This will become
more apparent during a more detailed discussion
of the motion mechanism below.

\begin{figure}[bt!]
\centering
 \includegraphics[width=\textwidth]{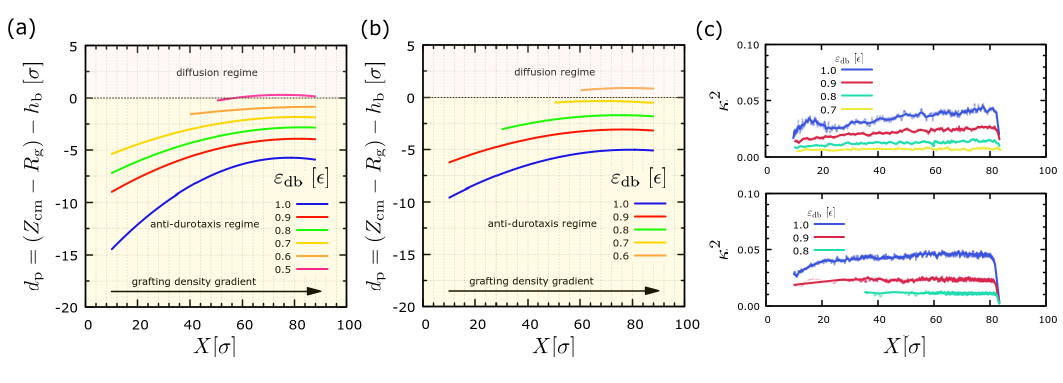}
    \caption{\label{fig:5} 
Penetration depth, $d_{\rm p}$, of the droplet into the substrate
as a function of the $X$ coordinate of the center-of-mass of
the droplet along the gradient (the direction of 
increasing $X$ is toward the regions with
higher grafting density, while anti-durotaxis motion is
towards smaller $X$ positions of the droplet) for different strength of
attraction $\varepsilon_{\rm db}$, as indicated.
Here, $\sigma_{\rm g, h}=0.6~\sigma^{-2}$,
$\sigma_{\rm g, l}=0.1~\sigma^{-2}$, $N_{\rm d}=10$
and $N_{\rm b}=50$ beads. Results for droplet size
$N=4000$ (a) and $N=16000$ beads (b) are shown.
(c) The dependence of the shape anisotropy
parameter $\kappa^{2}$ as a function of $X$ for droplets
with $N=4000$ (upper panel) and $N=16000$ beads (lower panel).
See text for details regarding the definitions of $d_{\rm p}$
and $\kappa^{2}$.
}
\end{figure}

Visual observations of successful anti-durotaxis
cases (for example, see the snapshots in
Fig.~\ref{fig:2}
and movie in the Supporting Information)
have led to the suspicion that the droplet 
gradually immerses into the brush as it moves
toward the regions of lower grafting density (softer regions),
since the droplet can be accommodated much easier
among the brush chains.
Hence, a logical next step is to attempt to characterize the
degree of penetration of the droplet into
the brush during the droplet motion, as part of the anti-durotaxis 
mechanism. Moreover, it would be
desirable to identify the boundaries between anti-durotaxis
motion and other cases, such as diffusive (non-anti-durotaxis or random)
droplet motion.
The penetration depth serves this purpose here and is defined as
$d_{\rm p} := (Z_{\rm cm} - R_{\rm g}) - h_{\rm b}$,
where $Z_{\rm cm}$ is the center-of-mass position
of the droplet in the $z$ direction, $R_{\rm g}$ its radius of gyration,
and $h_{\rm b}$ the height of the brush, which is 
identified by the inflection point of 
its density profile $d\rho^{2}/dz^2 = 0$ within the bin that
the $X$ position of the center-of-mass of the droplet belongs.
Hence, the penetration depth $d_{\rm p}$ expresses
the degree of the droplet immersion into the
brush, since $d_{\rm p}=-h_{\rm b}$ when 
$Z_{\rm cm}=R_{\rm g}$, \textit{i.e.}, the bottom of the droplet just touches the top of the brush. 
In turn, one can rescale the 
values of $d_{\rm p}$ according to $h_{\rm b}$, and
a value of zero would simply reflect the brush surface.
Figure~\ref{fig:5} presents our results for $d_{\rm p}$ for
different cases, which clearly show that the droplet
penetrates deeper into the brush as it moves to 
areas of lower grafting density 
(smaller $X$) during anti-durotaxis.
Moreover, the droplet immersion is deeper as 
$\varepsilon_{\rm db}$ increases. 
This initially correlates with a higher velocity
(Fig.~\ref{fig:3}) of the droplet, 
but further increase of $\varepsilon_{\rm db}$,
namely $\varepsilon_{\rm db}=\epsilon$, leads to
a slower anti-durotaxis motion, despite the
difference of the average velocity between
$\varepsilon_{\rm db}=0.9~\epsilon$ and
$\varepsilon_{\rm db}=\epsilon$ being rather small.
When the size of the droplet increases (for example,
see Fig.~\ref{fig:5}b, where plotted data are for
droplet with $N=16000$ beads), the conclusions remain the
same, but the drop in the penetration length during
anti-durotaxis is much smaller than in the case of droplet
with $N=4000$ beads, which reflects \blu{a weaker effect of the gradient and} a lower efficiency
of motion in the case of larger droplets (for example, see Fig.~\ref{fig:3}).
Overall, the data indicate that successful anti-durotaxis
motion is strongly related to the droplet penetration
into the brush.

As the droplet penetrates deeper into the brush and moves
toward softer areas (smaller grafting density) in successful 
anti-durotaxis experiments, it is relevant to investigate 
how the droplet shape changes during the motion.
Here, we measured the
shape anisotropy, $\kappa^{2}$,
of the droplet as a function of the
position $X$ of the droplet center-of-mass. 
In particular, $\kappa^2 = [b^2 + (3c^2/4)]/ {R^{4}_{\rm g}}$ with
$R_{\rm g} = \sqrt{\lambda^{2}_x + \lambda^{2}_y + \lambda^{2}_z}$,
and $\lambda^{2}_x, \lambda^{2}_y, \lambda^{2}_z$ the 
eigenvalues of the inertia tensor for the droplet beads.
$b$ is the asphericity defined as 
$\lambda^{2}_z - (\lambda^{2}_x + \lambda^{2}_y)/2$ 
and $c$ the acylindricity defined as
$\lambda^{2}_y - \lambda^{2}_x$.
The expectation is that $\kappa^2$ obtains values 
closer to zero for a spherical symmetry and unity
in the case of a cylindrical one. 
The results of Fig.~\ref{fig:5}c indicate that the droplet
initially transforms into a \blu{quasi} spherical-cap droplet as
it is deposited onto the substrate, which manifests itself
by the ``sharp'' increase in $\kappa^2$ and generally maintains
this symmetry throughout the anti-durotaxis motion with
$\kappa^2$ however remaining small throughout the motion.
Also, the droplets appear to have more spherical shapes
for smaller values of $\varepsilon_{\rm db}$.
Moreover, as the droplet moves to the regions
with lower grafting density, it gradually acquires
a slightly more spherical symmetry, due to the softness of
the substrate and the more available space
in these regions, despite, as we will see later,
the higher number of
interaction bead-pairs with the surrounding
brush polymer chains. Since these interactions are
at the substrate--droplet interface (no \blu{full} penetration
\blu{as the one defined in Fig.~\ref{fig:2}}),
they mainly minimize the energy of the
system through the droplet immersion into 
the brush rather than leading to changes in the elastic energy
of the droplet.

\begin{figure}[bt!]
\centering
 \includegraphics[width=\textwidth]{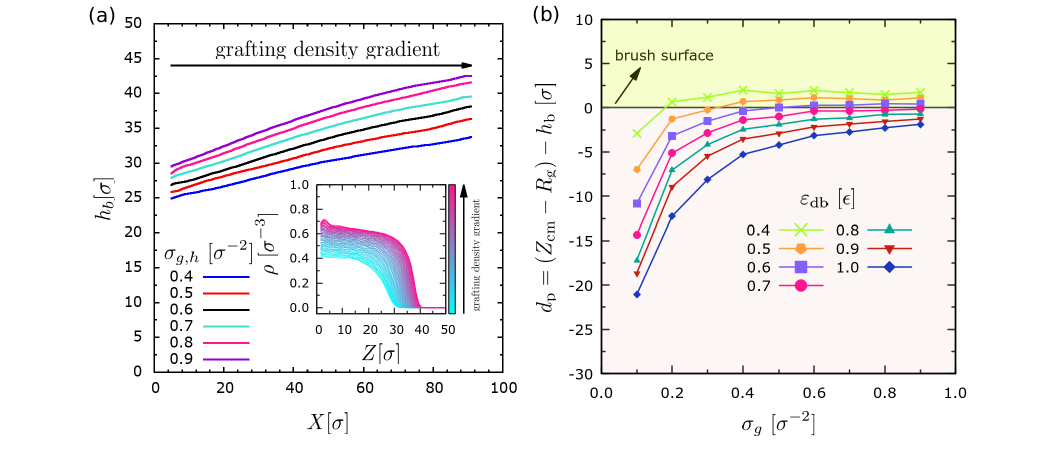}
\caption{\label{fig:6} (a) Brush height, $h_{\rm b}$, 
calculated by the inflection point of
the number-density profile, 
$d\rho^{2}/dz^2 = 0$,
along the $x$ direction at \blu{each} position $X$
for brush substrates with different 
gradients as defined by the different
$\sigma_{\rm g,h}$. In this case, 
the substrate is simulated without the droplet.
Also, increasing $X$ values correspond to areas
with larger grafting density.
Inset shows the number-density of the beads,
$\rho$, in  the direction normal to the brush 
substrate ($z$)
along the stiffness gradient (different color)
for the case $\sigma_{\rm g,h}=0.6~\sigma^{-2}$.
(b) Depth of penetration of the droplet,
$d_{\rm p}$, into substrates \textit{without} gradient of
different grafting density $\sigma_{\rm g}$,
as indicated. Data for different 
droplet--substrate attraction strength
$\varepsilon_{\rm db}$ are shown.
In the case of (a) and (b), data refer to
cases with $N_{\rm d}=10$, $N=4000$, and
$N_{\rm b}=50$ 
beads.
}
\end{figure}

In Fig.~\ref{fig:6}a, we closer examine the brush
properties without placing the droplet on the different
substrates,
thus providing further evidence for our previous arguments.
Our results indicate that the height of the brush,
$h_{\rm b}$, decreases toward the regions with
lower grafting density (smaller $X$). 
This trend does not depend on the particular
gradient in the grafting density (different values
of $\sigma_{\rm g,h}$ with $\sigma_{\rm g,l}=0.1~\sigma^{-2}$
remaining constant). Hence, the slope of the curves
in Fig.~\ref{fig:6}a is similar. 
However, larger values of the grafting density directly 
correlate with a larger height of the brush for a given
$X$ position according to the known expression
$h_{\rm b} \sim N_{\rm b} \sigma_{\rm g}^{1/3}$.\cite{Binder2012}
Examining the density profiles for a particular case
with a specific gradient of the grafting density
(inset of Fig.~\ref{fig:6}a),
we can also observe how the brush surface transforms from a
sharper to a wider interface as we examine
density profiles toward regions
of lower grafting density. Hence, not only the 
average height, $h_{\rm b}$, of the brush changes,
but, also, the structure of the interface, which sets
favorable conditions for the penetration of the droplet
into the substrate.
To complete the picture and better understand the above aspects, 
we have placed a droplet of $N=4000$ beads
onto brush substrates of different grafting density, $\sigma_{\rm g}$, 
and \textit{without} gradient in the grafting density
of the chains, and varied the
strength of the attraction between the droplet and the brush (Fig.~\ref{fig:6}b).
Then, we measured the penetration depth, $d_{\rm p}$.
First, we confirm that the droplet will penetrate deeper
into the brush when the grafting density becomes smaller.
From the point of view of  
\textit{in silico} nanoindentation experiments, 
a larger penetration depth 
reflects a softer substrate.\cite{Poma2019}
In the case of substrates with larger grafting density,
$d_{\rm p}$ decreases and rather reaches a plateau when
$\sigma_{\rm g,h}>0.4~\sigma^{-2}$. This suggests that 
smaller effects be expected in the droplet motion when the 
gradient becomes larger by setting
$\sigma_{\rm g,h}$, and this might explain the 
lower efficiency of anti-durotaxis motion for
larger gradients in the grafting density (Fig.~\ref{fig:2}).
Second, we observe that larger droplet--substrate attraction
also leads to a larger penetration depth. This effect seems
to be proportional to the attraction strength for a given
grafting density. Moreover, small values of $\varepsilon_{\rm db}$
and larger grafting density leads to situations where
the droplet practically levitates on top of the brush. 
These cases indicate a weak attraction of the droplet
to the substrate. One may assume that these
will not cause anti-durotaxis motion, since
this kind of motion is strongly driven by the
interfacial interactions between the droplet and the substrate,
which we will further expand upon below.

\begin{figure}[bt!]
    \centering
    \includegraphics[width=\columnwidth]{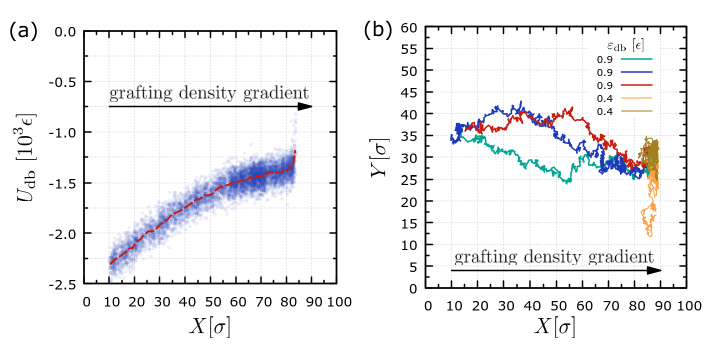}
    \caption{\label{fig:7} 
    (a) Interfacial energy between the droplet and 
    the substrate, $U_{\rm db}$, as a function of the
    position of the center-of-mass of the droplet, $X$,
    in the $x$ direction.
    (b) Typical trajectories of the center-of-mass positions
    $X$ and $Y$ of the droplet on the $x-y$ plane. 
    Trajectories for $\varepsilon_{\rm db}=0.9~\epsilon$
    refer to successful anti-durotaxis cases 
    with different initial conditions based on a
    different velocities distribution, while
    data for $\varepsilon_{\rm db}=0.4~\epsilon$
    correspond to trajectories of diffusive/random
    droplet motion.
    The droplet beads in each case have different 
    initial velocities.
    In both (a) and (b), $N=4000$, $N_{\rm d}=10$, 
    and $N_{\rm b}=50$ beads, 
    $\sigma_{\rm g,l}=0.1~\sigma$, 
    and $\sigma_{\rm g,h}=0.6~\sigma$.
    \blu{$\varepsilon_{\rm db}=0.9~\epsilon$ in panel (a).}
    }
    
\end{figure}

In previous studies,\cite{Theodorakis2017,Kajouri2023,Chang2015} 
it has been determined that the minimization of the interfacial
energy, $U_{\rm db}$, is the driving force for durotaxis motion,
which is confirmed for different substrate designs.\cite{Theodorakis2022}
Hence, it is also relevant for our study to examine how 
the droplet--brush interfacial energy varies during the anti-durotaxis
motion. Figure~\ref{fig:7} presents results of the
interfacial energy, $U_{\rm db}$, as a function of the
center-of-mass position of the droplet, $X$, for a 
typical anti-durotaxis case. The results confirm that 
the interfacial energy of the droplet decreases
as it moves from the higher grafting density
regions to the lower ones. As we have seen already above,
this also correlates with a deeper immersion of the droplet
into the brush, which indicates that the penetration of the
brush by the droplet sets the conditions for the energy minimization
of the system.
Furthermore, the energy profile is characterized by an initial
smaller slope in the decrease of the energy as a function of
the position $X$ of the center-of-mass of the
droplet, and then by a larger slope, which correlates 
well with the results of the penetration depth
(Fig.~\ref{fig:5}), $d_{\rm p}$,
and indicates that the driving force increases in 
the softer parts of the brush. In these softer
parts, we also observe that the driving force remains
constant until the completion of the anti-durotaxis
motion, as manifested by the constant gradient,
$F_{x} = \partial U_{\rm db} / \partial x$, which
is seen when approximately $X<60~\sigma$.
By monitoring various trajectories of successful
anti-durotaxis cases for a particular set of parameters (Fig.~\ref{fig:7}b),
we can clearly see that the motion of the droplet
as viewed from the top ($x-y$ plane of the brush) initially
appears more diffusive (random) and then much more
forward-moving, thus closely reflecting the observations
regarding the driving force in Fig.~\ref{fig:7}a.
For the sake of comparison, we also show cases of unsuccessful 
anti-durotaxis attempts for low $\varepsilon_{\rm db}$
values, where droplets show a random motion around
the very initial position $X$ of the droplet on the substrate.
This further
indicates that anti-durotaxis motion is not governed by
random fluctuations, but it is a result of the particular
choice of a set of parameters leading to a 
driving force, $F_x$, as in previous \textit{in silico}
experiments.\cite{Kajouri2023} In our case, the key set of
parameters is specifically the gradient in the grafting density
given the choice $\sigma_{\rm g,l}=0.1~\sigma^{-2}$ and
the choice of materials for the brush and the
droplet, which will eventually determine the strength 
of interaction between these two system components.

\begin{figure}[bt!]
\centering
 \includegraphics[width=\columnwidth]{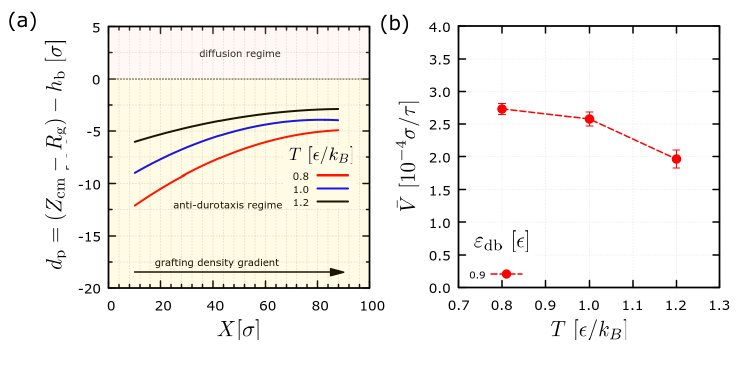}
\caption{\label{fig:8} (a) Penetration depth, $d_{\rm p}$, as a function
of the position, $X$, of the center-of-mass of the droplet for
different system temperature, $T$, as indicated.
(b) Average velocity of the droplet as a function of the temperature.
Here, $N=4000$, $N_{\rm d}=10$, and $N_{\rm b}=50$ beads.
$\varepsilon_{\rm db}=0.9~\epsilon$ and $\sigma_{\rm g,h}=0.6~\sigma^{-2}$. Data are based on an ensemble of 13
independent trajectories for each set of parameters to
obtain sufficient statistics.
}
\end{figure}

In the following, we explore the effect of the system
temperature on the anti-durotaxis motion. 
A case of highest motion efficiency is chosen
to facilitate the analysis and the
temperature of the system is varied. 
Overall, we find that anti-durotaxis motion will
take place for all of the three temperatures
studied here (one temperature lower and another greater than 
$\epsilon/k_B$). 
Moreover, we monitor the penetration depth 
and calculate the average velocity of the droplet center-of-mass, 
with results being presented in Fig.~\ref{fig:8}.
We find that the motion becomes more efficient (on average droplet moves faster)
in the case of the lowest temperature (Fig.~\ref{fig:8}b).
Moreover, we find that the more efficient motion correlates
well with a larger penetration depth of the droplet and
a larger slope of the depth as the droplet moves
toward the regions with smaller grafting density (Fig.~\ref{fig:8}a).
Hence, we may argue that the anti-durotaxis motion
be more efficient at lower temperatures, when
thermal fluctuations are less pronounced in the
system, and crucially at the droplet--substrate 
interface, which provides further evidence on
the underlying mechanism of the droplet motion, 
that is the minimization of the droplet--substrate interfacial
energy as the droplet is able to establish a larger number of
interaction contacts
with the brush during anti-durotaxis motion.
Note that the brush chains do not penetrate into
the droplet chains, and therefore, anti-dutoraxis is fully
controlled by the interfacial interactions, as
in the case of another brush-substrate design.\cite{Kajouri2023}
The increase of the temperature
generally leads to a decrease in the surface tension of the
droplet. However, temperature is also expeccted to
affect the substrate properties. Hence, the 
synergistic effect of the temperature can only accurately
be assessed by simulating
systems at different temperature, as is 
done in our study.

\begin{figure}[bt!]
\centering
 \includegraphics[width=\columnwidth]{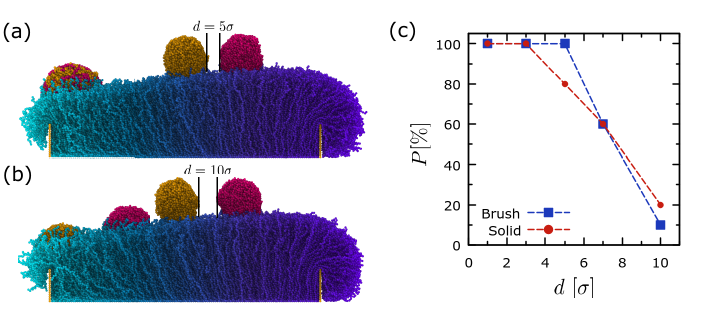}
\caption{\label{fig:9} Anti-durotaxis motion
of two droplets, which are initially placed
at a closer distance (\textit{i.e}, 5~$\sigma$)
and coalesce (a) and at a larger distance
(\textit{i.e.}, 10~$\sigma$) avoiding 
coalescence (b). 
In (a) and (b), the same snapshot has been used
in each case, since the substrate is only used 
for visualization purposes (see videos in 
Supporting Information for more representative examples.)
In both (a) and (b) the snapshots of the droplets
are taken at an initial time and at a later time
when both (a) or at least one (b) droplet has
reached the substrate end with the lowest grafting density.
(c) Probability that droplets
coalesce as a function of the initial distance
between the droplets, $d$. An ensemble of ten
independent trajectories has been considered for
the analysis. Here, $N_{\rm d}=10$,
$N_{\rm b}=50$, and $N=4000$ beads. 
$\varepsilon_{\rm db}=0.9~\epsilon$,
$\sigma_{\rm g,h}=0.6~\sigma^{-2}$, and
$\sigma_{\rm g,l}=0.1~\sigma^{-2}$.
\blu{The snapshot of the system was obtained using 
Ovito software.\cite{Stukowski2010} } 
}
\end{figure}

Although the study of
a single droplet on gradient substrates offers advantages
for \textit{in silico} experiments,
for example, isolating the various effects and carefully investigating
the droplet--substrate interactions,
multiple droplets on the same substrate are often used to
carry out the studies in real experiments.\cite{Style2013} 
This might be of benefit, for example, gathering 
statistics over a larger number of droplets within 
an individual experiment.
While the focus of this study is on the anti-durotaxis
motion of a single droplet, we also performed
\textit{in silico} experiments with two droplets
and carried out an ensemble of ten trajectories for
each case to explore the behavior of the system in such
a scenario. Again, the most efficient case was
chosen for the investigations with results presented in Fig.~\ref{fig:9}. 
Here, the main focus is placed on the role of the substrate
in the coalescence of the droplets, in other words
to probe whether the brush would favor the
droplet coalescence by acting as a ``bridge'' between
the droplets given the softness of the substrate, or
the brush chains would rather act as a ``barrier'' that
prevents the coalescence of the droplets.
To answer this question, we place two droplets
at different distances between each other onto two
different substrates, namely the brush substrate with
gradient of Fig.~\ref{fig:9}
and a smooth, unstructured substrate without gradient modeled by 
a 12--6 LJ potential assuming the same set of
interaction parameters. We find that the droplets will
first coalescence and then move together as a larger
droplet due to anti-durotaxis when the distance is small
enough in the case of the brush substrate. 
Moreover, the probability of coalescence depends on 
the distance $d$ between the droplets with the brush
substrate favoring coalescence over slightly larger distances
in comparison with the case where the droplets are placed
on the solid substrate.
In particular, for $d \leq 3~\sigma$, both substrates will lead to
droplet coalescence with 100\% probability. The solid
substrate then will provide droplet coalescence with
a smaller probability and after a distance $d \geq 6~\sigma$
the probability is more or less the same for both
substrate types. Since the potential cutoff
is $2.5~\sigma$, this might suggest that the
solid\blu{, smooth, unstructured} substrate not favor coalescence when the droplets
are not able to ``feel'' each other. In contrast, the brush
substrate gives 100\% probability of coalescence even 
when the distance between the droplets is $5~\sigma$, which
is twice the cutoff distance of the potential.
We might argue that the brush chains in this case
fill in the space between the droplets and act
in favor of coalescence. Further increase of the
initial distance between the droplets, $d$, however,
will weaken this effect and the behavior of the droplets
in terms of coalescence is the same for both the solid and
the brush substrate. 
Finally, we find that when coalescence between the
droplets is avoided, it will not
take place later during the anti-durotaxis motion 
of the droplets (Fig.~\ref{fig:9}b,
see also Supporting Information for the coalescence
and noncoalescence cases of Fig.~\ref{fig:9}).
It would be interesting to explore
systems with multiple droplets in the future and attempt
to further estimate the role of the brush chains as
mediators between the droplets or explore synergistic effects
that might be relevant in the anti-durotaxis motion
in systems of multiple and diverse droplets, 
but this clearly goes beyond the scope of the current study.

\section{CONCLUSIONS}
In this study, we have proposed and investigated
a new design of a brush substrate that is able
to lead to the self-sustained motion of droplets without
an external energy source. An important difference of
this substrate is that the droplet motion takes
place in the opposite direction of the stiffness 
gradient, hence the term anti-durotaxis that is coined
in our work. As such, to the best of our knowledge,
this is the first time that anti-durotaxis motion 
has been observed in simulations, since previous cases
were mainly concerned with durotaxis motion, that is
droplet motion in the direction of the stiffness 
gradient, from softer to stiffer regions of a substrate.\cite{Theodorakis2017,Chang2015,Kajouri2023}
Also, durotaxis motion is usually observed in 
experiments in the lab and biological 
systems,\cite{DuChez2019,Khang2015,Lo2000,Pham2016,Lazopoulos2008} while anti-durotaxis motion has only been
observed for a particular experimental setup.\cite{Style2013}

As in the case of durotaxis onto brush substrates\cite{Kajouri2023},
our analysis here confirms that the minimization of the
interfacial energy constitutes the driving force
that underpins the anti-durotaxis phenomenon. 
However, in the case of anti-durotaxis, this minimization
is caused by the gradual penetration of the droplet into the brush
as the droplet moves to the softer parts
of the substrate with lower grafting density. 
We have also conducted a parametric study in order
to gain further insights into the influence of the
various system parameters. We have concluded 
that soft brushes are in general more suitable
for anti-durotaxis, therefore fully flexible 
polymer chains and the lowest grafting density
on the one end of the substrate were always kept
$\sigma_{\rm g,l}=0.1~\sigma^{-2}$. Then, the 
two key parameters defining the probability of
success and the efficiency of the motion in 
terms of the average droplet velocity are the 
gradient in the grafting density and the
droplet--brush attraction strength. 
In particular, we find that large values of the
droplet--brush interaction strength (low surface energy) favor the 
anti-durotaxis motion, when full penetration of the
droplet is avoided, namely $\varepsilon_{\rm db}=0.9~\epsilon$
approximately. Then, moderate values of the grafting-density
gradient are preferable, in particular 
$\sigma_{\rm g,h}=0.6~\sigma^{-2}$.
This is due to the requirement of maintaining
the substrate softness that allows its penetration
by the droplet while maintaining a highest possible
gradient, \textit{i.e.}, a highest possible driving force.
As the droplet size
increases smaller gradients are performing better, but
the anti-durotaxis motion overall is less efficient
in the case of larger droplets. 
The temperature also plays an important role. We
find that a lower temperature will favor the
anti-durotaxis motion, which provides further evidence
supporting the minimization of the
droplet--substrate interfacial energy as the driving force
of anti-durotaxis. Finally, we
find that the viscosity has a minor
effect on the anti-durotaxis droplet motion, as 
it has been also observed in the case of the durotaxis motion
for a different brush--substrate design.\cite{Kajouri2023}

The key element of the new brush design 
is the gradient in the grafting density, and 
therefore it might be more applicable in the case of experiments,
since chemically same type of monomers can be used
for all chains. Hence, 
our findings might motivate further experimental
research in the area of self-sustained fluid motion
onto brush gradient substrates. 
This might be combined with experiments that
include population of droplets, which may
provide further understanding of the
underlying mechanisms of anti-durotaxis.
Here, we have also found that brush substrates
favor the coalescence of droplets
placed at a close distance in comparison with
smooth, unstructured, solid substrates.
\blu{Another aspect that may require further
investigation in the future is to assess possible effects of capillary wave-like fluctuations along the brush surface by simulating much larger
systems.}
We expect that future studies will provide
experimental evidence to the plethora of
possibilities unfolding regarding anti-durotaxis
motion. 
Also, we anticipate that this study already highlights 
new possibilities in the design of gradient substrates
and, as the first anti-durotaxis \textit{in silico} design,
holds important implications for various technological areas,
as it aims at forging understanding of the 
underlying mechanisms that underpin this and similar phenomena.

\begin{acknowledgement}
Authors thank Piotr Deuar, Rachid Bennacer, and Sergei Egorov for helpful discussions.
This research has been supported by the National Science Centre, Poland, under grant
No. 2019/35/B/ST3/03426. A. M. acknowledges support by COST (European Cooperation in
Science and Technology [See http://www.cost.eu and https://www.fni.bg] and its 
Bulgarian partner FNI/MON under KOST-11). 
We gratefully acknowledge Polish high-performance
computing infrastructure PLGrid
(HPC Centers: ACK Cyfronet AGH)
for providing computer facilities and 
support within computational
grant no. PLG/2022/015512.
\end{acknowledgement}

\begin{suppinfo}

M1*.mp4: Movie illustrates an example of
anti-durotaxis motion onto a gradient brush substrate. 
The parameters for the system are: $N=4000$, $N_{\rm d}=10$,
and $N_{\rm b}=50$ beads. $\sigma_{\rm g,h}=0.6~\sigma^{-2}$, 
$\sigma_{\rm g,l}=0.1~\sigma^{-2}$,
and $\varepsilon_{\rm db}=0.9~\epsilon$.

\noindent
M2*.mp4: Movie illustrates a different view of the system in M1*.mp4.

\noindent
M3*.mp4: Movie illustrates the case of Fig.~\ref{fig:9}a.

\noindent
M4*.mp4: Movie illustrates the case of Fig.~\ref{fig:9}b.
\end{suppinfo}

\bibliography{bib_brush}

\end{document}